# Diagnosing added value of convection-permitting regional models using precipitation event identification and tracking

Won Chang · Jiali Wang · Julian
Marohnic · V. Rao Kotamarthi ·
Elisabeth J. Moyer

**Abstract** Dynamical downscaling with high-resolution regional climate models may offer the possibility of realistically reproducing precipitation and weather events in climate simulations. As resolutions fall to order kilometers, the use of explicit rather than parametrized convection may offer even greater fidelity. However, these increased model resolutions both allow and require increasingly complex diagnostics for evaluating model fidelity. In this study we use a suite of dynamically downscaled simulations of the summertime U.S. (WRF driven by NCEP) with systematic variations in parameters and treatment of convection as a test case for evaluation of model precipitation. In particular, we use a novel rainstorm identification and tracking algorithm that allocates essentially all rainfall to individual precipitation events (Chang et al, 2016). This approach allows multiple insights, including that, at least in these runs, model wet bias is driven by excessive areal extent of precipitating events. Biases are time-dependent, producing excessive diurnal cycle amplitude. We show that this effect is produced not by new production of events but by excessive enlargement of long-lived precipitation events during daytime, and that in the

W. Chang
Department of Mathematical Sciences, University of Cincinnati, Cincinnati, Ohio.

J. Wang
Environmental Science Division, Argonne National Laboratory, Lemont, Illinois.

J. Marohnic
Center for Robust Decision Making on Climate and Energy Policy, University of Chicago, Chicago, Illinois.

V. Rao Kotamarthi
Environmental Science Division, Argonne National Laboratory, Lemont, Illinois.

E. J. Moyer
Department of the Geophysical Sciences, University of Chicago, Chicago, Illinois.
Tel.: +1-773-702-8101
Fax: +1-773-702-9505
E-mail: moyer@uchicago.edu

domain average, precipitation biases appear best represented as additive offsets. Of all model configurations evaluated, convection-permitting simulations most consistently reduced biases in precipitation event characteristics.

# 1 Introduction

The last decade has seen widespread use of dynamical downscaling via regional climate models (RCMs) as a means of producing detailed local simulations tied to global forcings. (See Xue et al, 2014, and references therein.) The assumption is that driving a high-resolution model (typically 4–25 km) with a coarser general circulation model (GCM) or reanalysis (∼100s km resolution) allows more realistically capturing processes that are inherently small in scale or controlled by small-scale land surface features. Dynamical downscaling seems especially promising for understanding potential future changes in precipitation characteristics (e.g. Chang et al, 2016), since convective cells are of ∼km spatial scale.

Although RCMs have been shown to improve representation of many aspects of climate, especially those involving the hydrological cycle (e.g. Wang et al, 2004), their output remains dependent on parameter choices, model configuration and the large-scale models that provide their boundary conditions (e.g. Racherla et al, 2012). When convection is parametrized, the choice of convective scheme has been shown to strongly affect precipitation characteristics (e.g. Giorgi and Mearns, 1999; Gochis et al, 2002; Xu and Small, 2002; Jankov et al, 2005; Awan et al, 2011; Pieri et al, 2015). Studies generally also agree that the microphysics scheme can impact precipitation (e.g. Jankov et al, 2005; Rajeevan et al, 2010; Bryan and Morrison, 2012; Awan et al, 2011; McMillen and Steenburgh, 2015; Gao et al, 2015; Pieri et al, 2015), though this conclusion is not universal (Kala et al, 2015). Other factors such as the planetary boundary layer and radiation schemes are of lesser importance (Xu and Small, 2002; Jankov et al, 2005; Awan et al, 2011; Gao et al, 2015). The sensitivity to convective scheme has motivated interest in higher-resolution (< 5 km) models that allow treating convection as resolved rather than parametrized. The assumption is again that this change will lead to more realistic simulations.

RCMs with parametrized convection commonly show biases in precipitation characteristics similar to those seen in coarser-resolution global simulations. GCMs tend to generate rain over too-large areas when compared to observations, with too-weak intensities: (the 'drizzle' problem) (e.g. Dai, 2006, and references therein). (A necessary consequence of this bias is that individual grid cells show too few dry days.) Convective precipitation in GCMs also tends to occur prematurely, presumably because of the difficulty of representing convective inhibition (e.g. Randall and Dazlich, 1991; Dai and Trenberth, 2004; Dai, 2006). RCMs with convective parametrizations similarly appear to generate rainfall over too large an areal extent (e.g. Davis et al, 2006a,b;

Chang et al, 2016) and, when convection dominates, with timing shifted earlier than the true diurnal precipitation cycle (Liang et al, 2004; Clark et al, 2007; Berenguer et al, 2012; Prein et al, 2013). The persistence of these biases in RCMs is unsurprising, as they also occur in global simulations at RCM-scale resolutions (e.g. Zhang et al, 2016, who show only modest sensitivity to resolution increases.) Too-large areal extents may be further exacerbated in RCMs by the fact that their total precipitation is frequently biased high. Examples of studies yielding high bias include Awan et al (2011); Done et al (2004); McMillen and Steenburgh (2015); Clark et al (2007); Berenguer et al (2012); Bryan and Morrison (2012); Gao et al (2015); Pieri et al (2015), among many others.

Convection-permitting simulations appear to somewhat alleviate the biases discussed above. The clearest effect is that removing the convective scheme generally delays the onset of convective precipitation, improving the match of the diurnal cycle to observations (e.g. Clark et al, 2009; Prein et al, 2013; Gao et al, 2017) (although Berenguer et al (2012) disagrees). Convection-permitting simulations also show changes consistent with reduced precipitation areal extent, though analyses are generally not conclusive, as they do not generally consider individual precipitation events. For example, Fosser et al (2015) show increased dry days at convection-permitting scale, consistent with reduced precipitation event size. Finally, these simulations also commonly generate lower total precipitation than counterparts with parametrized convection, reducing any initial wet bias. Examples include (Done et al, 2004; Pieri et al, 2015; Andrys et al, 2015; Prein et al, 2013). (Exceptions include Clark et al (2007) and Chan et al (2013), who show general increased wet bias, and Fosser et al (2015) who show an increase in winter.)

However, removing the convective scheme does not eliminate all precipitation biases. In particular, the few studies that treat areal extent in convection-permitting simulations continue to show too-large precipitating events. In a study of a single precipitation event (nested WRF with an inner grid of 1.33 km), McMillen and Steenburgh (2015) show that the model produces a rainstorm too large by 120-220% for high precipitation region ($\geq 10$ mm). More indirectly, Fosser et al (2015) see too-long durations of continuous rain, which the authors interpret as resulting from too-large precipitation events.

It is important to note that evaluating the effect of RCM parametrization and configuration choices (and the value of dynamical downscaling itself), is made difficult by multiple factors. 1) Regional climate models are inherently local, but the effects of parameter choices may differ by region. (e.g. Giorgi and Mearns, 1999; Liang et al, 2004). 2) The effects of parameter choices are not independent and may interact in complex ways (e.g. Jankov et al, 2005; Awan et al, 2011). 3) Sensitivity studies generally cover limited parameter space because of their computational demands, meaning effects may be confounded. For example, most studies do not independently consider horizontal resolution and the treatment of convection, but improved resolution alone does appear to affect precipitation amount (e.g. Bryan and Morrison, 2012). 4) Comparison studies usually analyze statistical properties of fields or timeseries at point locations, rather than the characteristics of individual events, which can hinder understanding of underlying physical drivers. Finally, 5) understanding the benefits of particular parameter choices is complicated by the problem of inherited bias. Multiple studies suggest that RCMs inherit bias from the large-scale models or data products used to provide their lateral boundary conditions (LBCs) (e.g. Yang and Wang, 2012; Bukovsky et al, 2013). If a given RCM parameter choice reduces that bias, it is difficult to know whether the change reflects an improvement in model physics or instead a degradation in model physics that inadvertently compensates for inaccurate boundary conditions.

These complications necessarily affect RCM precipitation studies, since the reanalyses often used as LBCs are typically wet-biased (e.g. Bosilovich et al, 2008). Downscaling analyses do not routinely compare RCM output with that of the underlying LBC, but studies have long suggested that LBC precipitation biases are heritable. For example, Warner and Hsu (2000) and Yang and Wang (2012) showed that precipitation varies strongly with choice of LBC, and Gao et al (2015) showed RCM bias ∼80% that of the underlying LBC. However, recent work suggests that wet bias is not the only cause of problems with precipitation characteristics in RCMs: in RCM simulations driven by an GCM with nearly unbiased total precipitation, Chang et al (2016) found that RCM precipitation events were nevertheless substantially too large, with their excess size compenstated by too-low intensities.

Analysis methods that identify and track individual precipitation events allow the richest examination of precitation characteristics in models. In principle, such methods can identify and distinguish between biases in event size, intensity, number, duration, and timing. In this study, we use the identification and tracking algorithm of Chang et al (2016) to develop new metrics for evaluating model fidelity at generating precipitation events that match observations. We use the algorithm to characterize model performance under a range of different model configurations, including convection-permitting simulations.

## 2 Model output and observational data

*Model simulations*

The bulk of this study involves a suite of model runs with systematic parameter variations generated in a preliminary investigation in support of the dynamical downscaling study of Wang and Kotamarthi (2014). All runs use as the regional model the Weather Research and Forecasting Model (WRF) version 3.3.1 (with the Advanced Research WRF dynamical core), and drive it with the NCEP Reanalysis II (NCEP) data product (Kanamitsu et al, 2002). Simulation domains span most of North America, but we restrict analysis to the continental United States (henceforth CONUS). Parameter testing runs are temporally short, covering the same 52-day summer period of June 1 to July 23, 2005, but we also compare to a similar NCEP-driven simulation run for 30 years, from 1980-2009. Runs are summarized in Table 1.

| Name | Res. | Conv. scheme | Microphys. scheme | Time |
|------|------|--------------|-------------------|------|
| CTRL | 12 km | GD | WSM6 | JJ 2005 |
| KF | 12 km | KF | WSM6 | JJ 2005 |
| Morrison | 12 km | GD | Morrison | JJ 2005 |
| 4 km | 4 km | none | WSM6 | JJ 2005 |
| *4 km Morrison* | *4 km* | *none* | *Morrison* | *JJ 2005* |
| *CTRL-30* | *12 km* | *GD* | *WSM6* | *1980-2009* |

**Table 1** WRF runs used in this study. Most analyses compare the first four (plain text), but we discuss the final two (italics) in specific contexts. For details see text.

Runs fall into two categories: 12 km resolution with parametrized convection and 4 km resolution with explicit convection. (The latter are performed on a slightly smaller domain, yielding total sampling of 600 × 516 grid points for the coarser runs and 1099 × 799 for the convection-permitting ones.) The bulk of the analysis in this work involves three runs at 12 km with differing convection and microphysics schemes, and one at 4 km (plain text in Table 1). Of the 12 km runs, the default (denoted *CTRL*) uses the Grell-Devenyi convective scheme (GD) (Grell and Dévényi, 2002) and WSM6 (WRF single-moment 6-class) microphysics (Hong and Lim, 2006); *KF* substitutes the Kain-Fritsch convective scheme (Kain, 2004) and *Morrison* the Morrison microphysics scheme (Morrison et al, 2005). The convection-permitting *4 km* and *4 km Morrison* runs use WSM6 and Morrison microphysics, respectively. All these simulations are initialized for one day (May 31th, 2005). *CTRL-30* has configuration identical to *CTRL* but is run for 30 years, with re-initialization each January 1. All runs use the Chen and Dudhia land surface model (Chen and Dudhia, 2001), the Yonsei University planetary boundary layer scheme (Noh et al, 2003), and the Rapid Radiative Transfer Model (Mlawer et al, 1997) for radiative forcing. See Wang and Kotamarthi (2014) for further details on WRF configuration and nudging by boundary conditions, and Wang et al (2015) for a full description of *CTRL-30*.

Prior studies that have explicitly compared the schemes tested here suggest that they strongly affect precipitation fields, though results are not always consistent, possibly because studies focus on geographically small areas. Multiple studies show wet bias in both KF and GD, but Awan et al (2011) finds GD wetter than KF (in WRF and a second RCM, over a domain in the Alps), whereas Yu et al (2011), Crétat and Pohl (2012), and Ratna et al (2014) find KF wetter than GD (in WRF, over domains in China and Southern Africa). Studies are in greater agreement that the Morrison scheme is an improvement over WSM6 (including the object-based Clark et al, 2014). Several analyses find that Morrison microphysics tends to reduce too-large areal extent of rainfall relative to both the single-moment WSM6 (Pieri et al, 2015) and the related double-moment WDM6 (McMillen and Steenburgh, 2015).

*Observational data and comparison protocol*

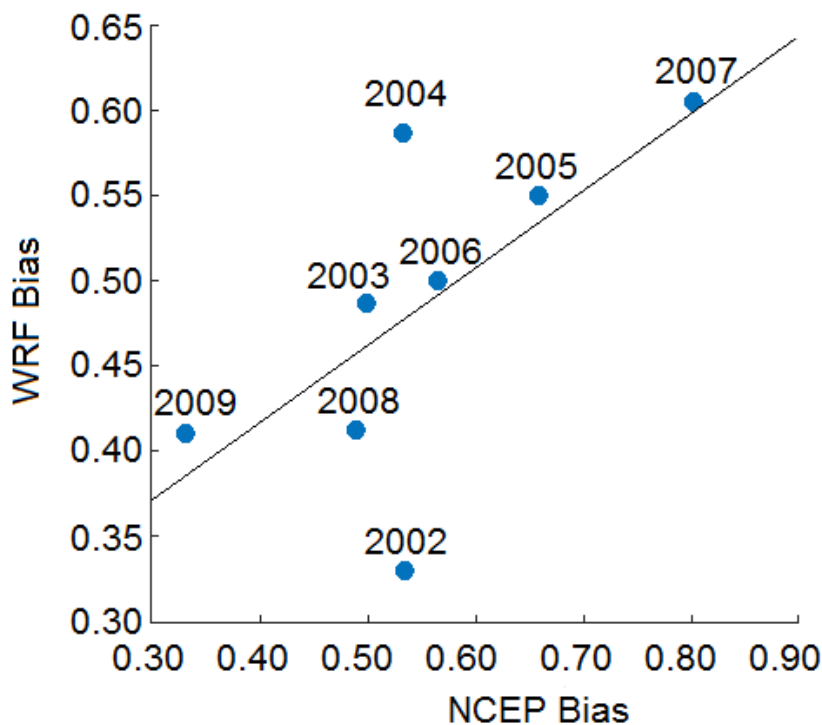

**Fig. 1** Comparison of biases in precipitation in NCEP-2 and in an RCM driven by NCEP-2. Biases are evaluated relative to Stage IV, for the 52-day June-July period and U.S. domain of this study. The strong interannual correlation suggests that precipitation bias in the RCM is largely inherited from its boundary conditions.

To evaluate model bias, we use the NCEP Stage IV precipitation data product, derived from a combination of radar and gauge data, gridded at 4 km resolution and reported at hourly intervals (Lin and Mitchell, 2005). Because Stage IV data is not reliable in the W. U.S. where radar stations are scarce, we restrict analysis when comparing to observations to the region east of longitude 114° W, where Stage IV data is valid. (That is, we use CONUS minus the W. Coast states, Nevada, and part of Idaho.) Storm identification and tracking are performed at native resolution, but in some analyses we upscale 4 km data to 12 km for consistency. Because some model runs were saved only at 6-hourly rather than 3-hourly intervals, all analyses use 6-hour cumulative precipitation.

*Large-scale boundary conditions*

It is important to note that the NCEP-2 reanalysis used to provide boundary conditions for these simulations carries a well-known wet bias, especially in summertime convective precipitation over land. Bosilovich et al (2008) compared reanalyses over 1979-2005 and reported that in NCEP-2, mean precip-itation over North America in July is 46% greater than in the Global Precipitation Climatology Project (GPCP) merged precipitation dataset (Adler et al, 2003). Over the timeframe and analysis domain of this study (described above), precipitation in NCEP-2 is 66% greater than in Stage IV.

To test whether the RCM is inheriting bias from the LBC, we use the 30-year *CTRL-30* simulation described above, and compare precipitation between *CTRL-30*, NCEP-2, and Stage IV in our 52-day period (June 1–July 23), for each year in which these datasets overlap (2002-2009). Both NCEP-2 and *CTRL-30* are too wet in all years, with mean biases of 55% and 49%, respectively. Biases show strong interannual variation that are indeed highly correlated, with a correlation coefficient of 0.48 (Fig. 1; the regression has slope 0.45 and intercept 0.24.) These results suggest that precipitation bias in our RCM runs is partially inherited from the driving reanalysis.

## 3 Methods

We compare the spatio-temporal characteristics of precipitation events in each WRF model run and in Stage IV by identifying and tracking individual precipitation events using the algorithm developed by Chang et al (2016), which relies on precipitation amount alone. The limited prior object-based studies of RCM output use relatively simple algorithms that group contiguous areas of precipitation above a threshold in a smoothed field (Done et al, 2004; Davis et al, 2006a,b; Clark et al, 2014). Because these methods must use a high threshold (0.3-4 mm/hour) to avoid chaining effects, they necessarily exclude a large portion of total precipitation. (In our datasets, a threshold 0.8 mm/hour as in Done et al (2004) would exclude 20-34% of all precipitation.) By using almost-connected-component clustering, the Chang et al (2016) algorithm allows decomposing precipitation into individual events using a much lower threshold and hence allows more complete analysis. (We use .03 mm/hour or 0.2 mm per 6-hour interval, which in our datasets excludes only 0.2-0.8% of precipitation.) Similar algorithms have been used in other contexts: see e.g. Baldwin et al (2005) or Murthy et al (2015).

Event identification in this algorithm is a two-step process: we first find for each timestep all contiguous regions of grid cells with precipitation exceeding the threshold value, and then group these regions into defined events using the modified almost-connected-component clustering described in Chang et al (2016). To track events, we connect identified clusters in consecutive time steps if they are morphologically similar and spatially close; the algorithm allows for both storm splits and mergers. The resulting "rainstorm objects" decompose the set of all precipitating grid cells into discrete events whose spatio-temporal evolution can be studied. As an example, see Figs. 2 and 3, which show precipitation fields and identified events for one six-hour interval on July 12, 2005, shortly after Hurricane Dennis made landfall.

In addition to analyzing characteristics of the individual constructed rainstorm objects (location, size, intensity, and duration), we can also decompose the total precipitation in each dataset into four factors based on the computed metrics for individual storms:

$$
\begin{aligned}
\text{Total Amount} = \text{Average Intensity} &\times \text{Size Factor} \\
&\times \text{Duration Factor} \times \text{Number of Rainstorms}.
\end{aligned}
\tag{1}
$$

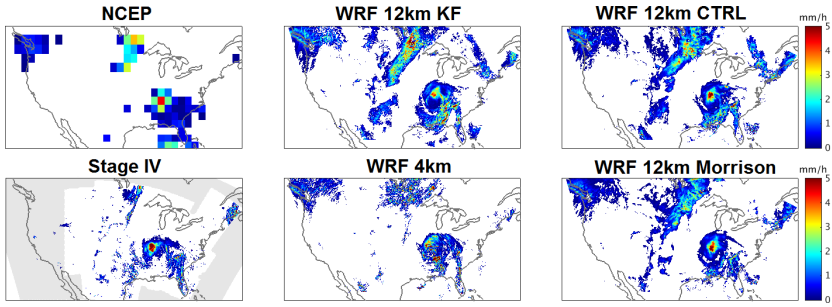

**Fig. 2** Comparison of precipitation in observations, reanalysis, and four reanalysis-driven WRF simulations; figure shows 6-hour cumulative precipitation for July 11, 12 PM-6 PM Time. Datasets show similar features, but areal extent of precipitation is too large in all 12-km simulations. The convection-permitting 4 km simulation more closely resembles observations. The cyclonic system in the SE is Hurricane Dennis, which made landfall in Florida on July 9 as a Category 3 storm. Grey in lower L panel marks areas where Stage IV data is incomplete or invalid.

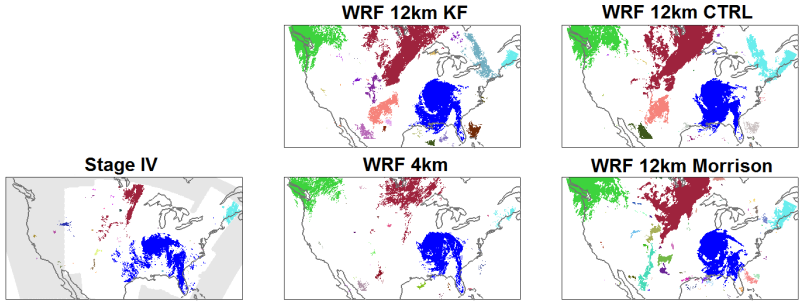

**Fig. 3** As in Fig. 2 but now with rainfall color-coded to mark individual events as identified by the storm identification and tracking algorithm. Hurricane Dennis is clearly identified in each case, but other precipitation events are differently labeled: for example, the V-shaped system in the Northeast is identified as one event in CTRL but two in K-F. Note that the algorithm allows a single event to include non-contiguous areas of precipitation.

(See Chang et al, 2016, Section 4b for details.) By comparing each dataset's decomposition to that in Stage IV, we can then quantify how bias in each factor contributes to the bias in total precipitation.

To visualize the spatial pattern of biases in each factor across the study area, we use spatial kernel smoothing (see Chang et al, 2016, Section 4c for details). For each grid cell location, we compute the spatial average of rainstorm metrics, weighted by kernel functions that give more weight to nearby storms and less to distant ones. We then map the ratios of these spatial averages between each model run and Stage IV at individual sites to show the spatial pattern of biases in each metric across the study area.

Finally, we examine structural differences within storms that cannot be captured by these metrics by comparing their radial profiles of precipitation intensity. That is, we plot the mean intensity as a function of distance from storm center, averaged across all individual storms. Storm centers are found by computing their 'centers of gravity'. (See Chang et al, 2016, Section S2 for details.) Because storms have very different sizes, we also construct normalized radial profiles. We first rescale individual storms, computing mean rescaled intensities (rescaled by dividing the maximum mean intensity for each storm) at different rescaled distances from the center of each storm (rescaled by dividing the maximum distance for each storm), and then construct an average profile across all individual storms. Note that because the rescaling is applied to individual storms, the resulting profile is different than it would be if created by simply rescaling the original mean profile.

## 4 Results: precipitation event characteristics

The example timestep of Figs. 2 and 3 shows that the identification and tracking algorithm captures precipitation events in an intuitive way and that all model cases reproduce the broad meteorological environment. However, precipitation events differ markedly across model cases in this example. Most strikingly, the parametrized-convection 12 km simulations all produce rain events that are substantially too large, while those in the explicit-convection 4 km simulation more closely resemble observations. In the example timestep, total areal extents of precipitating regions in *CTRL*, *KF*, and *Morrison* are 190, 180, and 160% larger than that in Stage IV. These size biases are larger than the corresponding wet biases (97, 120, and 40%, respectively), meaning that model intensities are too weak. By contrast, the explicit-convection *4 km* model run has similar areal extent as observations (-5.4%), and a smaller wet bias (23%) driven by enhanced intensity. Note that while the number of identified events in this timestep is large (16-35 across all model cases and Stage IV), the bulk of precipitation ($> 85\%$) from two major systems comprising 2-3 events: Hurricane Dennis (blue) and the central U.S. frontal system (brown/peach). These traits are broadly seen across all 209 timesteps in our simulations. For additional examples, see Supplementary Information Figs. S1-S3 and http://geosci.uchicago.edu/~moyer/Precipitation/movies/ for animations.

*Aggregate precipitation properties*

To summarize precipitation characteristics across all timesteps, we decompose total precipitation biases for each model case into factors for intensity, size, duration, and number of storms, as described in Section 3 (Table 4).

| Storm property | CTRL | KF | Morrison | 4 km | NCEP |
|---|---|---|---|---|---|
| Amount | 58 | 68 | 30 | 29 | 66 |
| Intensity | -13 | -21 | -21 | 20 | -17 |
| Size | 150 | 220 | 79 | 33 | - |
| Duration | -9 | -4.6 | -6.6 | -0.01 | - |
| Num. of storms | -19 | -42 | -1.9 | -20 | - |

**Table 2** Decomposed factors explaining precipitation bias for the model cases discussed in text and their LBC, expressed as % anomaly vs. Stage IV. All parametrized-convection runs have substantially too large precipitation event size; amount biases largely follow size biases but are less extreme. The explicit-convection *4 km* case appears more faithful to observations, deviating from its driving LBC. See Supplemental Information Table S1 for additional model cases.

Results are consistent with the example timestep shown in Figs. 2–3: the parametrized-convection 12 km runs produce precipitation events that are substantially too large and too weak in intensity, while the explicit-convection *4 km* run produces events more similar in size to observations and somewhat too strong in intensity. In the 12 km runs, changing the microphysics scheme has a larger effect than does changing the convection scheme. Using Morrison microphysics significantly reduces areal extent, consistent with results of prior studies (e.g. Clark et al, 2014); and reduces wet bias by nearly the same factor (∼50%). However, it remains unclear whether the reduced wet bias in both the *Morrison* and explicit-convection *4 km* runs is a purely positive attribute, as it means these runs deviate from the underlying reanalysis that drives them.

*Precipitation distribution within events*

It is important to check one potential factor that could confound size and intensity biases: whether the apparently larger rainstorm size in some runs is produced by 'haloes' of diffuse light precipitation that surround otherwise identical cores. Drizzle haloes would artificially inflate storm sizes, because the identification algorithm simply flags all pixels with precipitation greater than a low (0.2 mm per 6-hour interval) threshold value as part of a precipitation event, and does not consider precipitation distributions within that event. A commonly-used indirect test for this artifact involves successively repeating an analysis with different intensity thresholds. We conduct a more direct test by explicitly examining the mean radial profile of precipitation intensities across storms, computing mean and normalized (scaled) mean profiles as described in Section 3. The unscaled radial profiles (Fig. 4, left) show that distributions of precipitation within rain events do indeed differ across model cases. All parametrized-convection 12 km simulations have lower intensity at storm center and higher intensity in the tail than do the Stage IV observations or the convection-permitting *4 km* run. After normalization for size, the scaled radial profiles (Fig. 4, right) are remarkably similar in all cases. These results imply that the diagnosed differences in size and intensity across model cases are true structural differences and not drizzle artifacts. The results are robust across all storm sizes, and similarities extend to the tails of distributions. (See Supplementary Information Figs. S4–S7.)

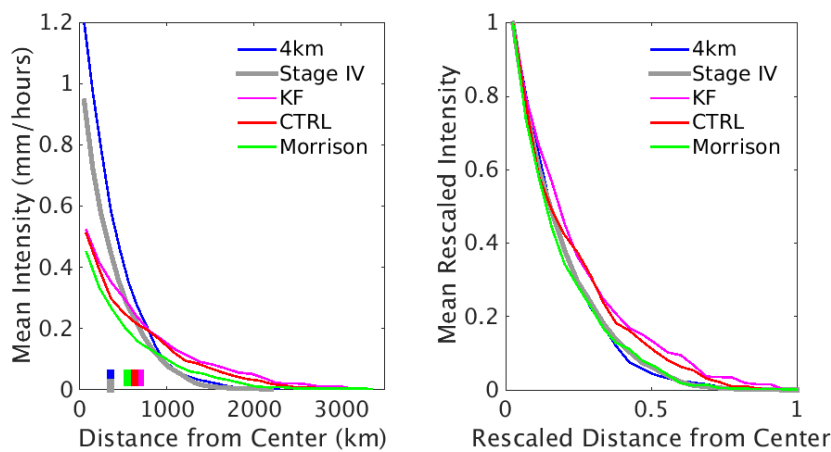

**Fig. 4** Original (left) and scaled (right) mean intensities as a function of distance from storm center, for events comprising the upper 80% of total precipitation amount. There is no evidence of a 'drizzle halo' effect that could artifically inflate derived storm sizes in some model runs: too-large storms in parametrized convection runs indeed have too-weak cores (left), and intensity distribution shapes are similar in all cases (right).

*Spatial comparison*

We also check for a second potential confounding issue, whether strong regional differences in precipitation event characteristics might make the domain-average analysis discussed so far misleading. Fig. 5 shows spatial patterns of biases in size, duration, and number for *CTRL*, *Morrison*, and *4 km*; see Supplementary Information Fig. S4 for *KF* and additional runs. Biases appear broadly similar over large portions of the domain but show some regional features. Most evidently, while all model runs are wet-biased, they also all show a band of too-dry territory in the Central U.S. associated with too-weak intensities. This feature was also seen in the GCM-driven simulations of Chang et al (2016); WRF fails to reproduce the sharp transition at the 100th meridian in precipitation amount and intensity. (However, reducing the nudging of model to LBC alleviates the bias; see Fig. S4.) Although intensity biases have spatial structure, the *difference* in intensity bias between the parametrized and explicit convection runs (-13-21% in *CTRL* and *Morrison* vs. +20% in *4 km*) is manifested fairly uniformly over the domain area.

The most noteworthy other spatial structure lies in size bias. In *CTRL*, size biases are relatively uniform over most of the domain area other than the Gulf Coast (precipitation events are too large as well as too weak). But in *Morrison*, which has reduced mean size bias over *CTRL*, the improvement occurs largely east of the 100th meridian, and size bias is actually positive in the Gulf Coast region. In *4 km*, spatial inhomogeneity in size bias is still greater, with a region of too-small precipitation events extending northwards from the Gulf Coast into the Great Plains. Biases in number and duration are relatively small and their spatial variations are likely not significant.

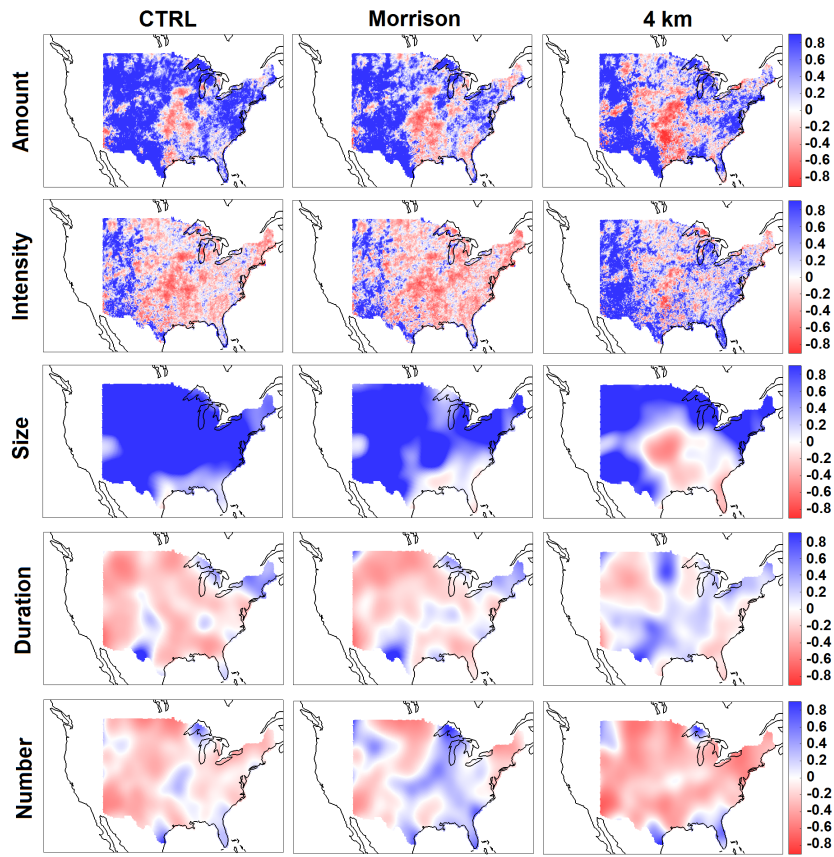

**Fig. 5** Fractional bias in precipitation characteristics for three model cases, evaluated vs. Stage IV data. Amount and intensity are evaluated at the pixel level but other factors are effectively spatially smoothed. Stippled patterns in amount and intensity biases derive from artifacts in Stage IV data associated with the edges of radar ranges. All model runs show a too-dry region in the Central U.S. that occurs because they fail to reproduce the sharp increase in intensity E. of the 100th meridian. Size biases show the most noteworthy other spatial inhomogeneities.

## 5 Results: precipitation distributions

Precipitation is a particularly challenging subject for statistical analysis given its strong dependence in both space and time. Precipitation distributions can provide a limited window into understanding the overall behavior of systems producing rainfall, but they necessarily lose or obscure some aspect of that spatio-temporal dependence. In this section we compare model output using three different approaches for describing precipitation distributions.

### Distribution by locations

Most analyses of precipitation distributions are 'location-based': they consider observed or simulated rainfall (integrated over some time interval) at fixed locations, either individual stations or grid cells. We will refer to these as precipitation 'incidences', distinct from the 'events' defined earlier, which cover multiple grid cells and propagate in time. Location-based analyses are often extended to larger areas by treating each station or grid cell identically and pooling all incidents when constructing a marginal distribution. Such analyses incorporate data from across an entire domain, but lose information on spatial relationships. The resulting distributions are heavily right-skewed for integration times from hourly to monthly. That is, the most frequent incidents are those of the lightest precipitation (e.g. Woolhiser and Pegram, 1979; Richardson, 1981; Stern and Coe, 1984). It is common to represent marginal distributions of precipitation (on wet days) as a two-parameter gamma distribution; the right skew then corresponds a small shape parameter (e.g. Husak et al, 2007). Too-large and too-weak precipitation events would manifest as an even heavier right skew, with too-frequent incidences of light precipitation and too few entirely dry days.

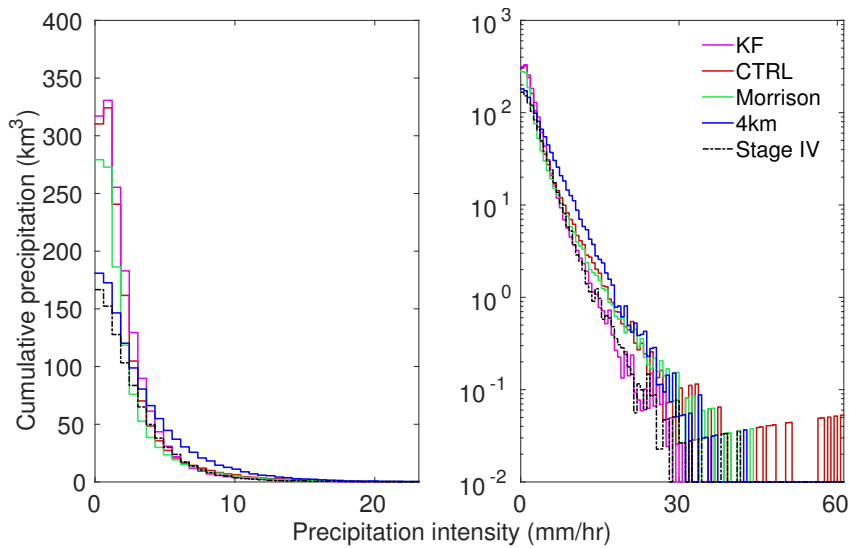

**Fig. 6** Location-based cumulative precipitation distributions for model runs and Stage IV observations, in linear (left) and log (right) scales, for all incidences exceeding threshold of 0.2mm/6 hour. Note that x-axis on left panel is truncated. All data are aggregated to 12 km grids and 6 hourly intervals ( >11M data points), but intensities on x-axis are stated in mm/hour for clarity. See Table S1 for totals across all intensities. On y-axis, 1 km$^3$ total precipitation is equivalent to ~60 incidents of mean intensity 20 mm/hour. All bins > 40 mm/hour represent single incidents. Distributions from parametrized- and explicit-convection runs differ in ways consistent with their differing amount and intensity biases.

Precipitation distributions in all our datasets are dominated by low-intensity rainfall, as expected (Fig. 6), but model distributions differ from that of Stage IV observations in ways consistent with their mean biases. (See Table 2 for biases.) The parametrized-convection runs *CTRL, KF* and *Morrison* have strong wet bias but too-weak mean intensity; in the distributions of Fig. 6, they show excess incidence of light precipitation. In these runs precipitation < 19.2 mm/6 hours (3.2 mm/hour) accounts for 96% of the total bias but comprises only 79-80% of total rainfall. The explicit-convection *4 km* run has a more moderate wet bias and too-strong mean intensity of similar scale; its distribution shows a relatively flat absolute bias over a wide range of intensities. The number of dry or sub-threshold incidents cannot be read off Fig. 6, but comprise 74-78% of the total for the parametrized-convection runs and 87% for both Stage IV observations and the explicit-convection *4 km* run.

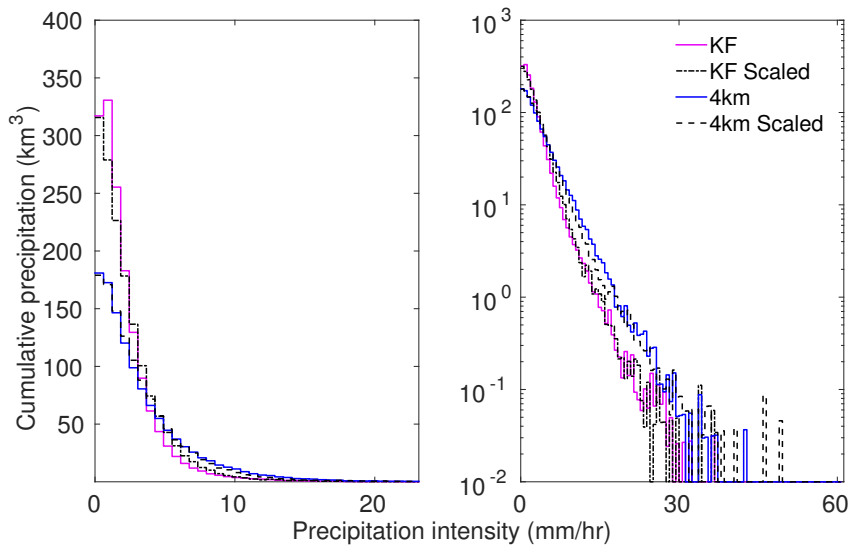

**Fig. 7** Cumulative precipitation distributions, as in Fig. 6, for the parametrized-convection *KF* and explicit-convection *4 km* runs (solid), compared to scaled versions of the distribution of Stage IV observations (dashed). (Stage IV is adjusted according to model amount and intensity biases, as descrbed in text.) This simple scaling appears to explain distributions well. See Supplementary Information Figs. S9–S12 for all runs.

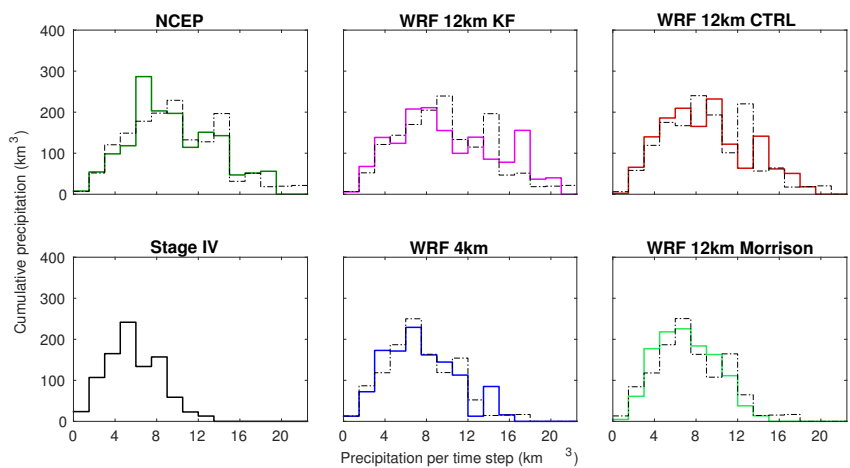

**Fig. 8** Cumulative precipitation distributions when rainfall is aggregated over the entire domain (i.e. 209 datapoints of 6-hourly rainfall in each case). Distributions for each model run and NCEP (colored solid lines) are compared to that of Stage IV scaled by to corresponding amount bias (dashed black lines). Most features of the distributions are well-explained by this simple scaling.

We then extend analysis beyond observing that model distributions are consistent with mean biases and test whether they are *fully* explained by those mean biases. That is, we compare each distribution to its 'null hypothesis', the simplest transformation of observations that we can generate using only two parameters, the model amount and intensity biases. For each model case, we adjust Stage IV observations as follows. We first multiply the amount of rainfall in each observed incident by the model intensity bias, then construct a cumulative precipitation distribution as in Fig. 6, then multiply the total rainfall in each bin by a factor $f =$ (amount bias)/(intensity bias). This procedure effectively assumes that model mismatch to observations is due to only two factors, a consistent under- or over-estimation of precipitation intensity and a mismatch in the number of wet vs. dry incidents.

The resulting 'null hypothesis' distributions match model results well in all cases, meaning that differences in location-based precipitation distributions are largely explained simply by the mean biases shown in Table 2. The parametrized-convection *KF* and explicit-convection *4 km* match extremely well at all intensities (Fig. 7, and see Supplementary Information Figs. S9–S12 for all runs). The *CTRL* and *Morrison* runs match well for the upper 97% of precipitation but show heavy high-intensity tails (incidences with precipitation > 60 mm/6 hours). The limitations of this dataset make it difficult however to assess the statistical significance of tail behavior due to the small sample size.

*Distribution using spatially aggregated precipitation*

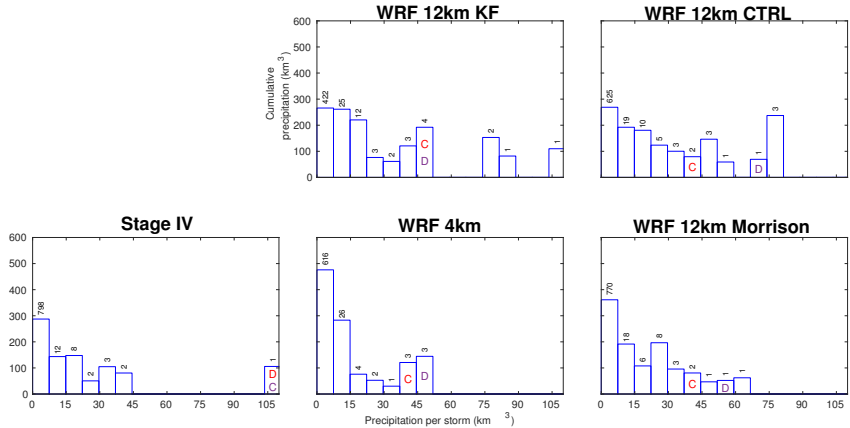

**Fig. 9** Event-based precipitation distributions classed by individual storm precipitation amount (i.e. total rainfall over the duration of the storm). Numbers above each bar give the number of individual storms in each size bin. (We exclude the smallest storms that collectively comprise 0.1% of total rainfall.) Labels 'C' or 'D' on a bin indicate the largest storm identified as part of Hurricanes Cindy or Dennis; the label position within a bar shows its rank relative to other events in that amount class. In Stage IV, Cindy and Dennis are confounded and identified as a single outlier event labeled 'DC' here.

An alternative approach to describing precipitation distributions, that retains the temporal information of which incidents have occurred in the same time step is to aggregate precipitation spatially across the domain. We therefore assess the distributions of total domain precipitation for the 209 six-hour timesteps of our model simulations, and seek to determine whether this formulation encompasses novel information for diagnosing model performance. After domain averaging, the cumulative precipitation distributions appear relatively symmetric in both observations and models, as expected due to the central limit theorem (Fig. 8; see also Supplementary Information Figs. S13–S14). Also as expected, the wet-biased model simulations show broader distributions than do observations, with higher maximum domain-averaged precipitation.

As before we compare to a 'null hypothesis', though a simpler one than in the previous example, since intensity bias is likely not relevant when aggregating precipitation across a scale much larger than that of individual events. We therefore simply scale the timeseries of domain-average Stage IV precipitation by each model's amount bias and then evalute the distribution. Most features of model distributions are indeed well-explained by this simple scaling (Fig. 8), suggesting that they contain little novel information.

*Distribution by events*

Finally, we examine distributions that make use of our identication and tracking algorithm, which breaks precipitation fields into discrete events. Because this approach takes into account spatio-temporal correlations in precipitation fields, it may provide information on model behavior beyond that already encoded in mean biases. Fig. 9 shows cumulative precipitation distributions by event size for each of our datasets (with size meaning total precipitation over the lifetime of an event); see also Supplementary Information Table S3. We show cumulative precipitation since in all runs the number count is dominated by tiny and short-lived events that contribute negligibly to total rainfall: across runs, 0.5% of precipitation is produced by the bottom 34-51% of all events, with median amounts $< 0.02$ km$^3$ and durations 12 hours. The cumulative distributions remain right skewed in all cases, i.e. the smaller events produce the largest fraction of total rainfall, but the runs show differences that cannot be explained with any simple scaling.

One striking feature is that the parametrized-convection runs with WSM6 microphysics (*CTRL* and *KF*) produce excessive contribution from a few individual large events. ('Large' here again refers to total precipitation amount rather than areal extent, though the two are correlated in all runs to greater than 0.97.) In Stage IV observations, Hurricanes Cindy and Dennis are the dominant precipitation events. The two are conflated by the identification and tracking algorithm, but their sum is more than double the next largest event. Cindy and Dennis are quite similar in all model runs (see Supplementary Information Table S3), but *CTRL* and *KF* also create multiple precipitation events substantially larger than either named hurricane. In *KF*, the largest non-hurricane event exceeds Cindy and Dennis combined. These large events contribute substantially more to total precipitation in *CTRL* and *KF* than would be expected from model biases: the top five non-hurricane events exceed expectations by 24 and 32%, respectively. Use of Morrison microphysics and explicit convection both remove the production of extremely large events: in *Morrison* and *4 km*, top five events actually underproduce expectations by -11 and -12%. The effects appear roughly additive, and combining them in *4 km Morrison* deepens the underproduction of large events (Supplementary Information Table S3 and Fig. S15). These results are robust when considering either a fixed number of storms or quantiles in the high tail of the event-based distribution.

A second clear feature of the event-based distributions is that, at least in these runs, the use of explicit convection produces excessive contribution from small events. From mean model bias alone, we would expect that events in the smallest amount class would contribute *less* to total precipitation in *4 km* than in Stage IV; instead, they dominate the *4 km* distribution and drive the bulk of total precipitation bias. The effect persists regardless of which microphysics scheme is used (Supplementary Information Fig. S15 for *4 km Morrison* comparison). The discrepancy largely stems from of a population of anomalously high-intensity ($> 1.5$ mm/hour over the lifetime of the storm) but short-duration ($\sim$1 day) and spatially small events originating from the Gulf Coast, that are seen only under explicit convection (Supplementary Information Fig. S16). See also Fig. S1 for an example showing the phenomenon clearly. This misrepresentation of Gulf Coast convection is presumably the source of the negative size bias seen in this region in both explicit-convection runs (Figs. 5 and S8).

While we do not have a clear explanation of the underlying physical processes in models that drive these biases, they demonstrate the complexity of precipitation spatio-temporal characteristics and suggest that simple diagnostics are not sufficient to evaluate models. The results also provide a caution that while explicit convection appears in many ways to improve model fidelity, it may also introduce new and more subtle biases.

## 6 Results: precipitation timing

*Relationship of the diurnal cycle to precipitation events*

Biases in precipitation timing are a well-known issue for many models, and the use of explicit convection is often assumed to improve the representation of the diurnal cycle. The particular model runs used here are not well-designed for studying precipitation timing, since output is analyzed only at relatively coarse 6-hourly temporal aggregation and the length of the runs is short. But the identification and tracking algorithm provides a useful tool for analyzing precipitation timing, because it allows addressing questions such as: to what extent the diurnal cycle in precipitation results from enhanced production or persistence of events at certain times of day? What processes drive biases in the diurnal cycle? We therefore seek to use our runs to help identify useful diagnostics of timing issues.

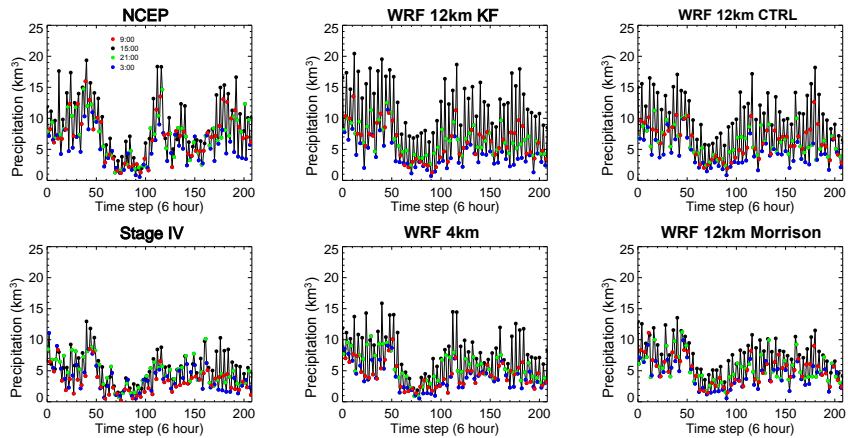

**Fig. 10** Domain-average precipitation per 6-hour timestep in observations and model runs. Colors indicate center of timestep, in Mountain Daylight Time (UTC - 6 hours). The domain extends over three time zones, but the 6-hourly aggregation means that we cannot divide data to show equivalent local times. All datasets exhibit diurnal cycles with expected afternoon maximum, but model afternoon peaks are too large and too regular. No model captures the anomaly associated with hurricanes on timesteps ∼130–170, where observed evening precipitation (green) often matches or exceeds the afternoon peak (black). (Cindy makes landfall on 7/6 at timestep 130 and Dennis on 7/10 at timestep 159.) Note that while the hurricanes are individually large events, they do not produce significant positive anomalies in domain-aggregated precipitation in any dataset.

All datasets analyzed here (Stage IV observations, NCEP reanalysis, and WRF model runs) show clear diurnal cycles in domain-average precipitation, with peak rainfall as expected in the afternoon (Fig. 10). In the model runs, however, the afternoon peaks are both too large and too regular. This effect is strongest for model cases with parametrized convection and WSM6 microphysics (*CTRL* and *KF*); the mean amplitude is moderated in cases of either Morrison microphysics or explicit convection, and only the explicit convection runs generate realistic day-to-day variations in amplitude. Models also show an earlier nighttime minimum than do observations (3 AM timestep vs. 9 AM in Stage IV); the effect is again worst in *CTRL* and *KF*. All models fail to capture one feature associated with Hurricanes Cindy and Dennis that is potentially significant, that on multiple days just after their landfalls, the observed diurnal peak broadens, i.e. precipitation remains high into the evenings.

Analysis of precipitation events shows however that the timing of event production and dissipation has little bearing on these biases, or on the diurnal cycle in general. Neither the number of events present at any given time (Fig. 11) nor the number of new event initiations (Supplementary Information Fig. S17) show variations that would drive an afternoon rainfall peak. The number of events present in each timestep is in fact anti-correlated with total rainfall, weakly for observations (correlation -0.2) and in model runs, more strongly the larger the size bias (up to -0.41 for *KF*, see Supplementary Information

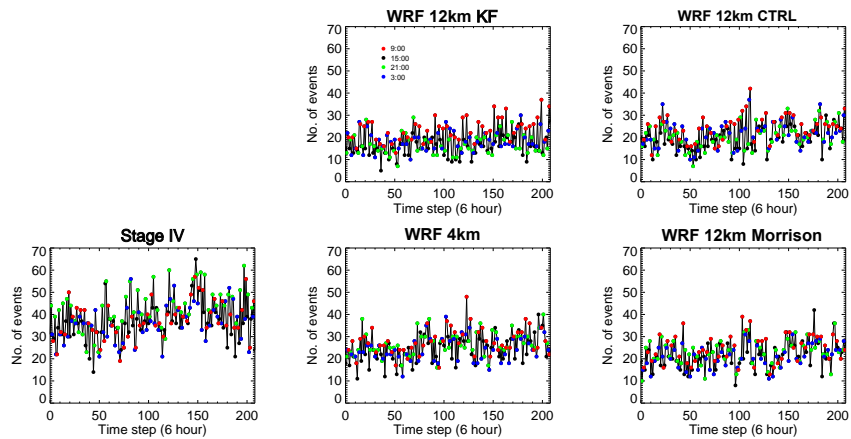

**Fig. 11** Number of precipitation events present in each 6-hour timestep in observations and model runs. Colors indicate center of timestep as in Fig. 10. For number of events *initiated* in each timestep, see Supplementary Information Fig. S18. Variations in event numbers do not drive the diurnal precipitation cycle in either observations or models. Instead, domain-aggregated precipitation is anti-correlated with event numbers (correlation coefficients -0.2 to -0.4; see Fig. S18.)

Fig. S18). The anti-correlation is in fact likely a consequence of this size bias, especially when combined the coarse temporal resolution: as the area of rainfall expands, nearby precipitation incidences are more likely to be identified by the algorithm as a single event.

Note also that the lifetimes of identified precipitation events are typically longer than a diurnal cycle. Mean duration, when events are weighted by precipitation amount, is over 2 days for the model runs evaluated here and over 3 for Stage IV observations (Table S1). While some previous studies using identification and tracking algorithms have found significantly shorter durations, their algorithms relied on higher thresholds and are therefore more likely to relabel a moving region of precipitation as a new event (e.g., Clark et al, 2007). Even our analysis here likely underestimates durations relative to what would be meteorologically intuitive, as the 6-hour aggregation results in more 'dropped' identifications and therefore shorter durations than in the study of Chang et al (2016) that used 3-hourly precipitation. The implication is that the diurnal cycle must result not from changes in event numbers but from time-dependent fluctuations in rainfall over the lifetime of individual events.

*The diurnal cycle and precipitation biases*

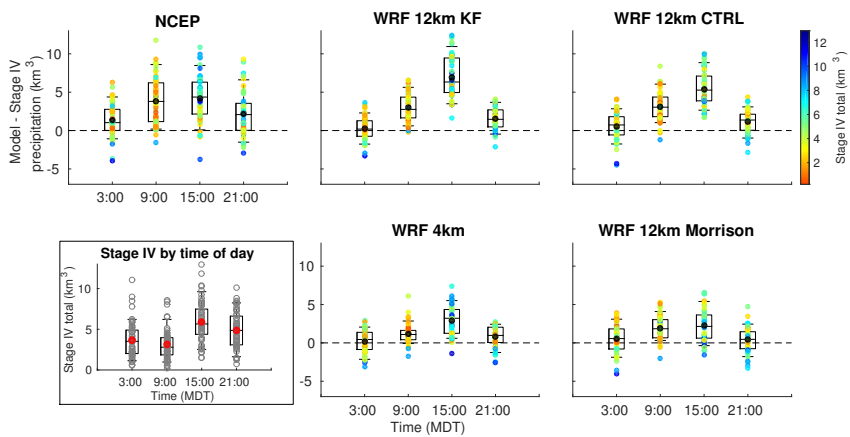

**Fig. 12** Bias in diurnal cycle: absolute bias in domain-aggregated precipitation by time of day (Mountain Daylight Time) for all model runs and their LBC (5 panels other than bottom left), and for comparison, diurnal cycle in Stage IV observed precipitation. X-axis labels mark center of 6-hour time intervals. In all panels, dots show means and box plots show the median, 25th and 75th percentiles, and 9th and 91st percentiles of each distribution. Color code in the bias plots indicates the total observed precipitation in each time step. Blue color in low outliers are cases where Stage IV is atypically high and models do not follow. All downscaled model runs show an amplified diurnal cycle, though using explicit rather than parametrized convection appears to moderate this effect. NCEP and *Morrison* also run have disproportionate morning bias that broadens the rainfall peak.

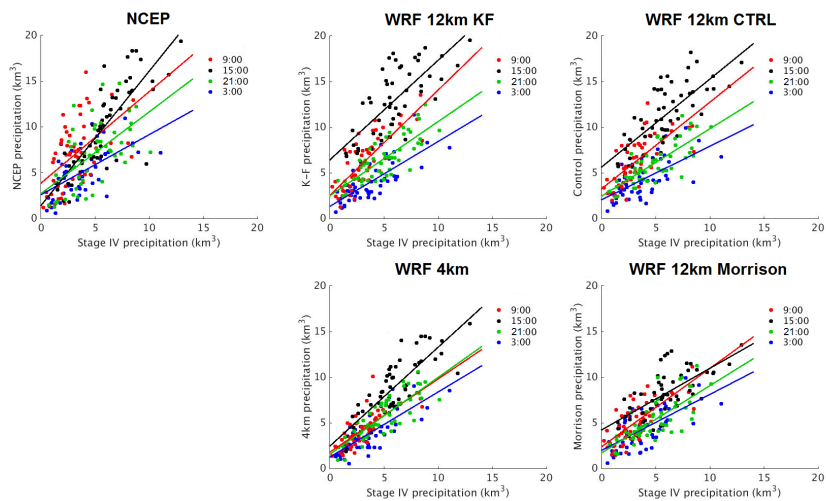

**Fig. 13** Domain-averaged model vs. observed (Stage IV) precipitation for all model cases and NCEP, by 6-hour time interval. Color code shows the four time intervals, in Mountain Standard Time. Lines show linear fits for each subset of the data. In regional model cases, larger afternoon biases (black) appear driven primarily by a change in intercept rather than a change in slope.

This over-production of daytime precipitation is robust even in those parts of the domain where the diurnal cycle has a different phase. The models analyzed here therefore amplify the diurnal cycle in the East, where observed precipitation peaks in the afternoon, but dampen it in the Central U.S. (W. of 93 W), where observed precipitation peaks at night. (See Supplementary Information Figs. S20 to S22 compare all-domain and regional diurnal precipitation cycles.) As previously seen (Table 2), mean bias is smallest in model cases with either explicit convection (*4 km*) or Morrison microphysics (*Morrison*). These cases have similar mean biases (Fig. 12) but different temporal patterns of bias, with the WSM6 microphysics producing a stronger afternoon peak. As before, adding the *4 km Morrison* model case allows comparing across all parameter combinations (Fig. S23). The effects appear approximately additive, with little interaction.

The complexity of these biases raises deeper questions about how model precipitation bias should best be described to capture the physical processes responsible. One approach for diagnosing the origin of biases is to consider their temporal variations. In Section 2, we applied that approach to year-to-year variations to assess whether precipitation biases in regional models were inherited. We can also apply the approach to day-to-day variations to assess whether bias should be more properly considered an absolute (additive) rather than a fractional (multiplicative) effect. In either representation, mean biases would vary over the diurnal cycle (Table S4), so insight requires examining not only the mean bias but its variations. Fig. 12 in fact already suggests an answer: we show absolute rather than fractional biases here because this representation produces a more compact relationship across model days.

The suggestion of an additive bias is more clearly seen in a scatterplot of domain-averaged model vs. observed (Stage IV) precipitation (Fig. 13, which shows precipitation in each time interval for the 52 days of this study). Linear fits for the separate time intervals are roughly parallel, supporting the ideal that the time-dependent biases are best seen as time-variable absolute off-sets in precipitation. The conclusion of an additive bias is still stronger when considering daily precipitation (Fig. 14). With the diurnal cycle removed, precipitation variations in all model cases are tightly correlated with those in observed precipitation (correlation coefficient over 0.82) and the relationship has slope close to 1 (range 0.88–1.07). Model bias in daily mean precipitation therefore appears well-described as an additive rather than multiplicative effect. It is worth noting that this conclusion, while robust, is strongest for the explicit-convection *4 km* case, which has the highest correlation (0.98) and the slope closest to 1 (0.97). The reduced mean bias that results from the use of explicit convection appears manifested primarily as a lowering of an additive offset in precipitation.

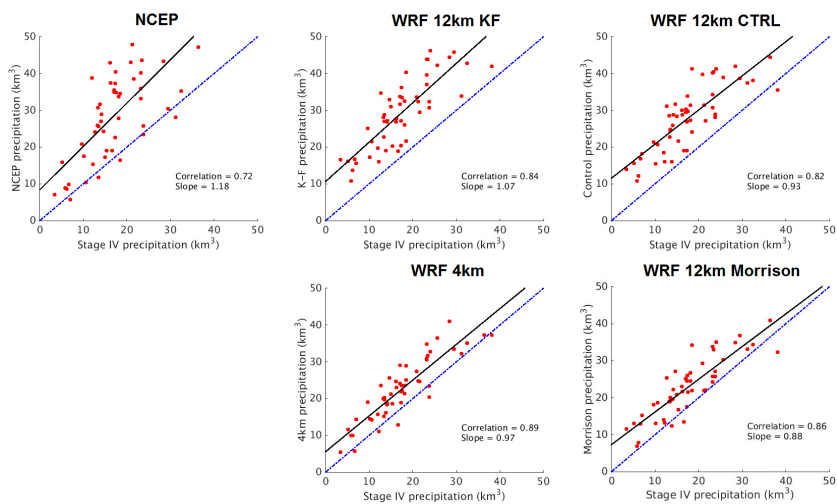

**Fig. 14** Domain-averaged model vs. observed (Stage IV) precipitation for all model cases and NCEP, by 24-hour day to remove the diurnal cycle. Lines show linear fits. Correlations are strong and slope close to 1 for all regional model cases, suggesting that bias in daily precipitation is best described as an additive effect. For an analogous figure that does not average across the diurnal cycle, see Supplementary Information Fig. S24.

# 7 Discussion and Conclusions

As regional climate modeling increases in resolution, simulations have become able to capture complex and potentially realistic aspects of precipitation events. These qualities in turn require new methods for evaluation of model fidelity. Precipitation is a particularly challenging variable to characterize statistically, given its non-Gaussianity and strong spatio-temporal dependences. The most appropriate means of describing precipitation characteristics is through identifying and tracking individual precipitation events, and we show here that use of even extremely simple identification and tracking algorithms makes possible useful new diagnostics of model performance. Although the suite of model runs we use here are is necessarily limited, the analysis shows the type of physical insights that can be derived from this approach.

Many useful diagnostics of course do not require event identification and tracking, including mean amount and intensity biases. In the model cases analyzed here, these mean biases seem to some extent to reflect those in their driving local boundary condition, NCEP, which is too wet but too low in intensity; all model cases studied here are also too wet, and most too low in intensity; the parametrized-convection runs with WSM6 microphysics have mean biases nearly identical to those of NCEP. When biases are considered separately by timestep, NCEP shows an exaggerated diurnal cycle and all model cases share this feature, most strongly in the case of the parametrized-convection runs with WSM6 microphysics. Of the model cases studied, those with explicit convection (*4 km* and *4 km Morrison*) deviate the furthest from their LBC.

It is important to recognize however that many seemingly promising diagnostics actually contain limited additional information. Location-based cumulative precipitation distributions are a popular diagnostic of model performance, but in all model cases here, distributions can be largely reproduced using only observations and model mean amount and intensity biases. Similarly, distributions of domain-average precipitation appear well-explained simply by model mean amount biases.

The identification and tracking algorithm allows additional insight that would not be possible otherwise. In this analysis, event-based diagnostics show that model overall wet biases are driven by anomalies not in the number or duration but in the size of individual events, and that model excess daytime precipitation results not from event generation or dissipation but from fluctuations of the size of long-lived events. That is, model rainstorms become too large in daytime. Finally, in cases with parametrized convection, the contribution of individual events to total precipitation is weighted too heavily to large events, while in cases with explicit convection, it is weighted too heavily to small events.

This analysis provides several consistent lessons about the effects of different model configuration choices. One persistently recurring result is that in cases with parametrized convection, the choice of convective scheme appears less important than the choice of microphysics scheme. This conclusion holds when considering overall amount bias, size bias, and the diurnal cycle. These results are contrary to the findings of some other studies (e.g. Pieri et al, 2015; Jankov et al, 2005), and may of course be a function of the particular choices tested. A more limited conclusion may simply be that the WSM6 scheme is particularly problematic.

A second recurring result across multiple evaluation metrics is that simulations with explicit convection appear more faithful to observations than those with parametrized convection. The finding of increased realism is consistent with many previous studies (e.g. Andrys et al, 2015; Done et al, 2004), but we can now demonstrate it in additional dimensions. In these runs, use of explicit convection produces not only reduced mean wet bias and a less inflated diurnal cycle, but smaller individual events more consistent with observations. The maximally realistic scenario is that combining explicit convection with Morrison microphysics; the effects appear roughly additive. It is however important to recognize that explicit convection is not a panacea for all model issues, and in these runs it is also associated with some new and complex regional biases.

The underlying causes of the precipitation biases here (or in any study) are not well understood. Why do models produce convective features over too large an area in daytime, and why does explicit convection partially mitigate that bias? Why do biases appear as absolute offsets? More broadly, we do not even know whether a the appropriate criterion for validating a regional climate model really should be matching observations, if that model has been driven by unrealistic boundary conditions. The goal of climate model evaluation is not simply to identify which simulations most closely match observations but to provide a pathway for fundamentally understanding model physical processes. Our identification and tracking algorithm may seem somewhat problematic for this purpose, since the event definitions are not themselves deeply rooted in physics. But any discrepancies flagged by event-based model diagnostics are nevertheless real, and can both suggest both physical causes and guide efforts at improvement.

**Acknowledgements** The authors thank Matthew Huber, Andreas Prein, and the participants in the 2016 GEWEX Convection-Permitting Climate Modeling Workshop for many helpful comments and suggestions. Christopher Callahan assisted with preparation of figures and diurnal cycle analysis. This work was conducted as part of the Research Network for Statistical Methods for Atmospheric and Oceanic Sciences (STATMOS), supported by NSF awards 1106862, 1106974, and 1107046, and the Center for Robust Decision-making on Climate and Energy Policy (RDCEP), supported by the NSF Decision Making under Uncertainty program award 0951576. Additional support was provided by the University of Cincinnati TAFT Research Center and computing resources were provided by the University of Chicago Research Computing Center.